\documentclass[a4paper,11pt]{article}
\usepackage{graphicx} 
\usepackage{tikz} 
\usepackage{placeins}
\usepackage{jinstpub} 
\usepackage{subcaption}
\usepackage{lineno}
\usepackage{amsmath}
\usepackage{multirow}
\usepackage{upgreek}
\usepackage{makecell}
\usepackage{booktabs}
\usepackage{threeparttable}
\usepackage{comment}
\usepackage{xcolor}
\usepackage{float}
\usepackage{pgfplots}
\usepackage{CJKutf8}
\usepackage{soul}
\usepackage{hyperref} 
\usepackage[numbers,sort&compress]{natbib}
\title{Simulation Based Characterization of Deconvolution-Based PMT Waveform Reconstruction Under Large Charge Dynamic Range and Varying Scintillation Time Profiles}

\author[a]{Xingyi Lin,}
\author[a]{Jinghuan Xu,}
\author[a,1]{Yongbo Huang, \note{Corresponding author.}}
\author[a,2]{Jingzhe Tang, \note{Corresponding author.}}
\author[a]{Tianying Xiao,}
\author[a]{and Yingke Li,}
\affiliation[a]{School of Physical Science and Technology, Guangxi University, Nanning 530004, China.}

\emailAdd{huangyb@gxu.edu.cn}
\emailAdd{2207301115@st.gxu.edu.cn}

\abstract{Photomultiplier tubes (PMTs) are widely used as photon sensors for neutrino and dark matter detection. Accurate charge and time information extracted from PMT waveforms is crucial for event reconstruction. An algorithm based on deconvolution technology was proposed and applied to the reconstruction of PMT waveforms. This study further investigated the reliability of the deconvolution algorithm when handling a large charge dynamic range (0-200 photoelectrons), varying scintillation time profiles, and muon-induced large signals. Monte Carlo data confirmed that the deconvolution algorithm exhibits relatively stable reconstruction performance: under the simulation conditions described in this paper (including a noise level of 0.1~PE, single photoelectron charge resolution of 30\%, 1~GHz sampling rate, 1000~ns window, three undershoot configurations, and eight scintillation time profiles), the residual non-linearity of charge reconstruction is controlled to approximately 1\% over the range of 0 to 200 photoelectrons, and the algorithm is capable of handling muon-induced large signals. The reconstruction performance depends on adequate baseline recovery; a waveform window that is too short relative to the undershoot tail leads to degraded reconstruction quality, which can be mitigated by extending the sampling window.}


\begin{document}
\begin{CJK}{UTF8}{gbsn}
\maketitle
\flushbottom

\section{Introduction}
\label{sec:intro}

PMTs are widely used photosensitive devices that play a critical role in neutrino and dark matter experiments~\cite{SNO:1999crp,Super-Kamiokande:2002weg,KamLAND:2004mhv,Borexino:2008gab,DayaBay:2016ggj,Cheng:2017usi,DarkSide:2018bpj,AKERIB2020163047}. In recent years, with the development of fast analog-to-digital conversion (FADC) technology, several particle physics experiments have achieved high-fidelity waveform acquisition across thousands of PMT channels, providing a data foundation for offline reconstruction. The primary goal of PMT waveform reconstruction is to extract high-precision charge and timing information from the signals, which are essential for event energy reconstruction and vertex determination~\cite{Li:2021oos,Yang:2022din,Qian:2021vnh,Zhang:2024okq,Zhang:2025dhu}.

In response to different experimental scenarios and PMT signal characteristics, researchers have proposed a variety of waveform reconstruction methods, including the simple charge integration method, CR-(RC)$^4$ method, waveform fitting, as well as approaches based on deconvolution techniques and machine learning~\cite{Huang:2017abb,Jiang:2024wph,Xu:2021whl,Wang:2024qxq}. Among these, the deconvolution-based reconstruction method was first introduced by the Daya Bay experiment. It is capable of effectively processing complex PMT waveforms that contain components such as undershoot and reflections, and successfully supported the direct measurement of electronic non-linear effects in the Daya Bay experiment~\cite{DayaBay:2019fje}. This method has since been further developed and applied in subsequent experiments, such as JUNO~\cite{Zhang:2019rfl,Grassi:2018pxk}. Optimized by incorporating the short-time Fourier transform (STFT)~\cite{Tang:2024jfs}, this approach has also demonstrated strong performance in the separation and identification of pile-up signals. Before proceeding, it is worth briefly comparing the deconvolution method used in this study with other PMT waveform reconstruction techniques mentioned in the literature. The simple charge integration method and the CR-(RC)$^4$ method are computationally efficient but are sensitive to baseline fluctuations and waveform shape features such as undershoot. Waveform fitting methods can achieve high accuracy but are often computationally expensive and rely on accurate waveform models; in addition, they may suffer from fitting quality issues. Machine learning based methods have shown promising performance in pile-up identification and photon counting, but they typically require large training datasets and lack interpretability. The deconvolution method used in this study offers a balance: it effectively removes waveform shape features (including undershoot and reflections) with moderate computational cost, and its parameters can be determined from calibration data without extensive tuning. Given that this paper focuses primarily on investigating the performance of the deconvolution method in several different application scenarios, a full quantitative comparison among these methods is beyond the scope of this study but is an important direction for future work.

In recent years, the development and operation of large-scale liquid scintillator detectors, such as the JUNO experiment, have introduced new requirements for the performance of PMT waveform reconstruction algorithms. These include, in particular, the stability of reconstruction performance under different scintillation time profiles (i.e., for various liquid scintillator formulations or when the scintillator is excited by different particle types), as well as the adaptability to operation across a large dynamic range in terms of charge. The former concern arises from the coupling between the scintillation time profile and the PMT waveform characteristics, which can affect reconstruction performance. For example, applying the simple charge integration method or the CR-(RC)$^4$ method to PMT waveforms that exhibit features such as undershoot can lead to biases in the reconstructed charge. This effect can propagate to event energy reconstruction, resulting in additional non-linearity in the energy response. Therefore, it is necessary to evaluate the stability of PMT waveform reconstruction algorithms under different scintillation time profiles. On the other hand, in large-scale liquid scintillator detectors with high light yield, PMTs usually operate over a wide charge dynamic range when detecting signals from different energy ranges. For instance, in the JUNO experiment, the operational range of the 20-inch PMTs for reactor neutrino detection is approximately 0 to several tens of photoelectrons (PEs), depending largely on the relative position between the energy deposition point and the PMT. For higher-energy particles such as atmospheric neutrinos and cosmic-ray muons, a single 20-inch PMT in JUNO may receive hundreds or even thousands of photoelectrons. Moreover, if the PMT waveform exhibits significant undershoot, the reconstruction of such large signals poses an even greater challenge.

In this study, building upon our previous research on deconvolution-based PMT waveform reconstruction, we further investigate the feasibility of applying this method under different scintillation time profiles and within large dynamic ranges in terms of charge. The effect of signal undershoot is also discussed, and the performance of the deconvolution algorithm is characterized across various application scenarios. The details of this work are as follows. Section~\ref{sec:simulation_deconvolution} introduces the simulated PMT waveform samples used in this study and briefly reviews the deconvolution algorithm. In Section~\ref{sec:performance}, the performance of the deconvolution algorithm is evaluated under different scintillation time profiles, and its adaptability to operation across a large charge dynamic range is examined. Considering the potential impact of event energy deposition patterns and the corresponding baseline recovery of waveforms on the application of the algorithm, this section separately analyzes PMT waveform reconstruction for point-like and track-like energy deposition events, and discusses the corresponding reconstruction performance. Finally, a summary is provided in Section~\ref{sec:summary}. It should be emphasized that the simulation parameters used in this study (such as the 0.1 PE noise level, 30\% SPE charge resolution, 1~GHz sampling rate, and 1000~ns sampling window; more details in Section~\ref{sec:simulation_deconvolution}) are motivated by the typical configurations of large liquid scintillator experiments, including JUNO. However, the present work is not intended as a full-detector simulation for any specific experiment. Instead, we aim to evaluate the deconvolution algorithm under generic yet realistic conditions.

\section{PMT waveform simulation and deconvolution reconstruction method}
\label{sec:simulation_deconvolution}

This section describes the generation of simulated waveforms used for the analysis and briefly reviews the deconvolution-based waveform reconstruction method.

\subsection{PMT waveform simulation}
\label{sec:simulation}

\begin{figure}[H]
    \centering
    \begin{subfigure}[t]{0.48\textwidth}
        \centering
        \includegraphics[width=\textwidth]{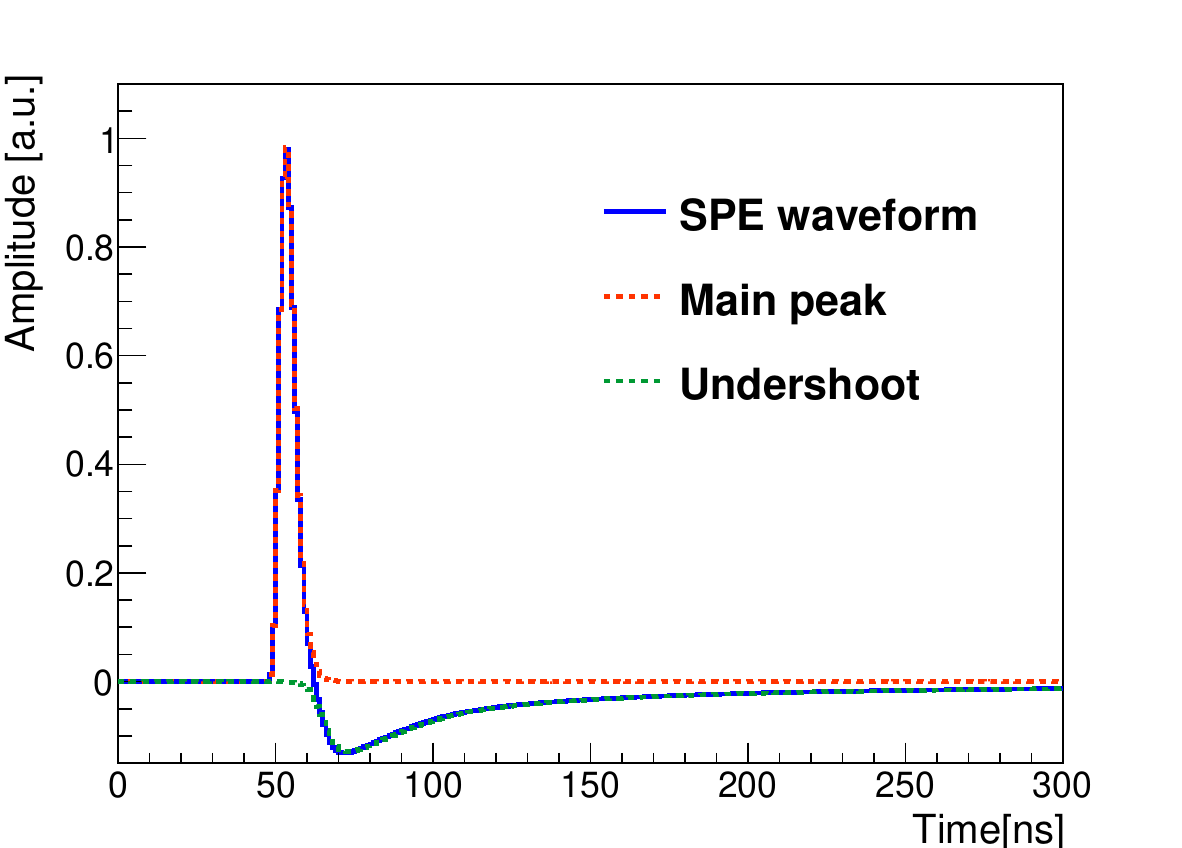}
        \caption{}
        \label{fig:spe_template}
    \end{subfigure}
    \hfill
    \begin{subfigure}[t]{0.48\textwidth}
        \centering
        \includegraphics[width=\textwidth]{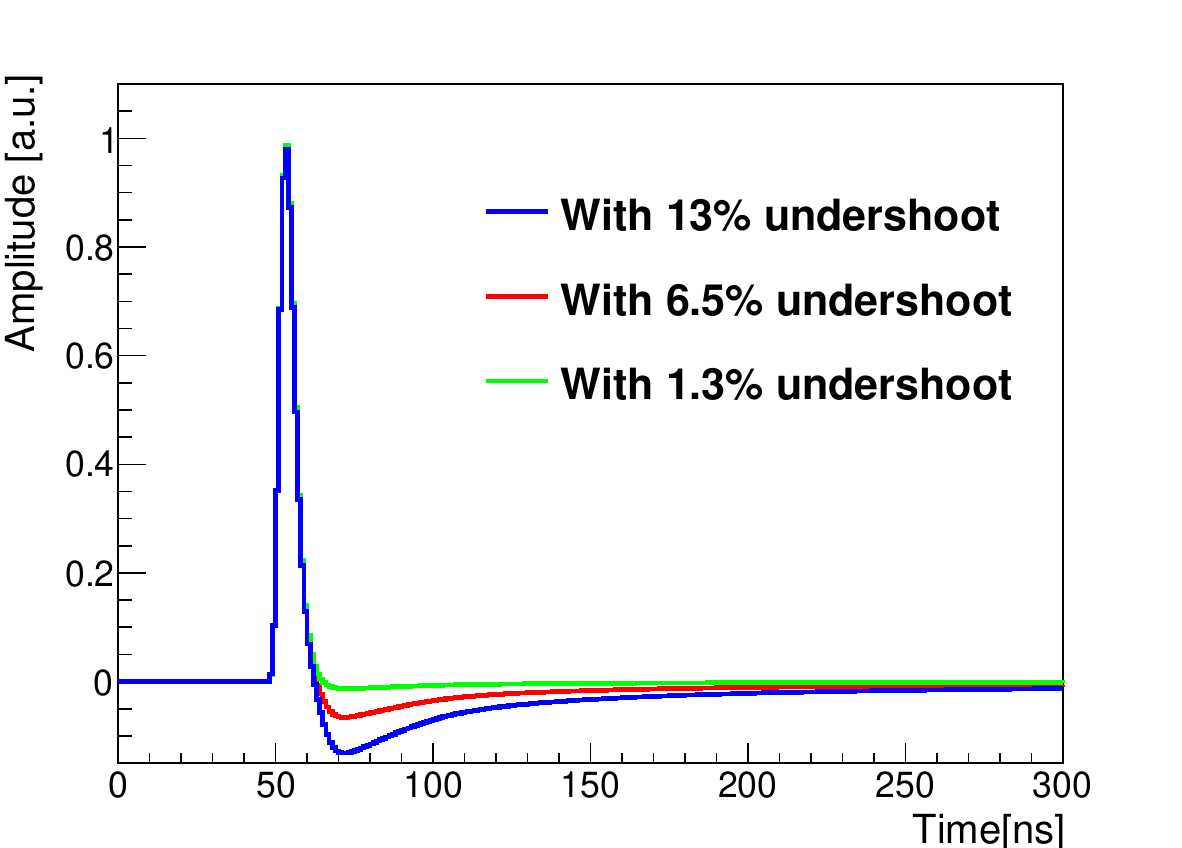}
        \caption{}
        \label{fig:spe_template_overshoot_comp}
    \end{subfigure}
    \caption{(a) A typical SPE waveform template. It consists of two components: the main peak and the undershoot. (b) SPE waveforms with different configurations of undershoots. To provide a clearer illustration of the waveform characteristics, only the segment from 0 to 300~ns is shown in the figure, while the full waveform length is 1000~ns. For all three undershoot configurations, the baseline of the SPE waveforms is fully restored by the end of the 1000~ns time window.}
    \label{fig:spe_waveform}
\end{figure}

In this study, simulated PMT waveforms are generated using the parameterization method described in~\cite{Jetter:2012xp}, with the relevant parameters obtained from~\cite{Huang:2017abb,Tang:2024jfs}. This parameterization follows the single photoelectron (SPE) waveform model developed for the Daya Bay experiment, which is the most detailed and publicly available PMT waveform model in the literature and includes waveform shape parameters and their fluctuations derived from data fitting. It should be noted that the present simulation does not include non-linear effects such as PMT saturation or electronic non-linearity. These effects are real and important in practical experiments; however, their characterization typically relies on dedicated calibration measurements~\cite{Wu:2022gdk,Yang:2019vog}. The focus of this work is on the waveform reconstruction algorithm itself, i.e., accurately recovering the charge encoded in a given PMT waveform, independent of whether the waveform already contains saturation-induced distortions. By excluding experiment-specific non-linearities, we aim to isolate and evaluate the intrinsic performance of the deconvolution algorithm. In real experiments, once the reconstructed charge is obtained, it can be further corrected using calibration-derived non-linear response curves to recover the true charge corresponding to the photons received by the PMT. A typical SPE waveform template is shown in Fig.~\ref{fig:spe_template}, which primarily consists of the main peak\footnote{To facilitate general discussion and comparison, the amplitude of the main peak is normalized to 1.} and an undershoot component. For simplicity, signal reflection is assumed to be absent. Reflections primarily originate from impedance mismatches in the backend circuitry. With careful impedance matching in the design phase, reflections can be largely suppressed. Moreover, as shown in~\cite{Huang:2017abb,Grassi:2018pxk,Tang:2024jfs}, the deconvolution method is insensitive to waveform shape features; including reflections would not affect the main conclusions of this study regarding undershoot and baseline recovery. Therefore, we omit reflections to keep the analysis focused. Furthermore, if the waveform does not return to baseline by the end of the time window, it may lead to spectral leakage—an effect that is particularly pronounced in large-charge waveforms with significant undershoot. To evaluate and compare the impact of this effect on large-charge waveform reconstruction, we adjust the undershoot amplitude and perform waveform simulation and reconstruction analysis using three types of SPE waveform templates, as shown in Fig.~\ref{fig:spe_template_overshoot_comp}. In these three templates, the ratios of the undershoot amplitude to the main peak amplitude are approximately 13\%, 6.5\%, and 1.3\%, respectively. The three undershoot amplitudes are motivated by experimental measurements. As reported in~\cite{Luo:2016ddb}, PMT undershoot originates from the discharge of the decoupling capacitor of the HV-signal decoupler and the capacitor in the HV divider. In the early stage of the Daya Bay experiment, without dedicated optimization of these components, the undershoot was approximately 13\% of the main peak amplitude. Through subsequent design optimizations, e.g., for JUNO, the undershoot can be reduced to about 1\%. The intermediate value of 6.5\% is also included for a systematic study. These three values thus represent poorly optimized, moderately optimized, and well-optimized conditions, respectively.

\begin{figure}[H]
\centering 
\includegraphics[width=1\textwidth]{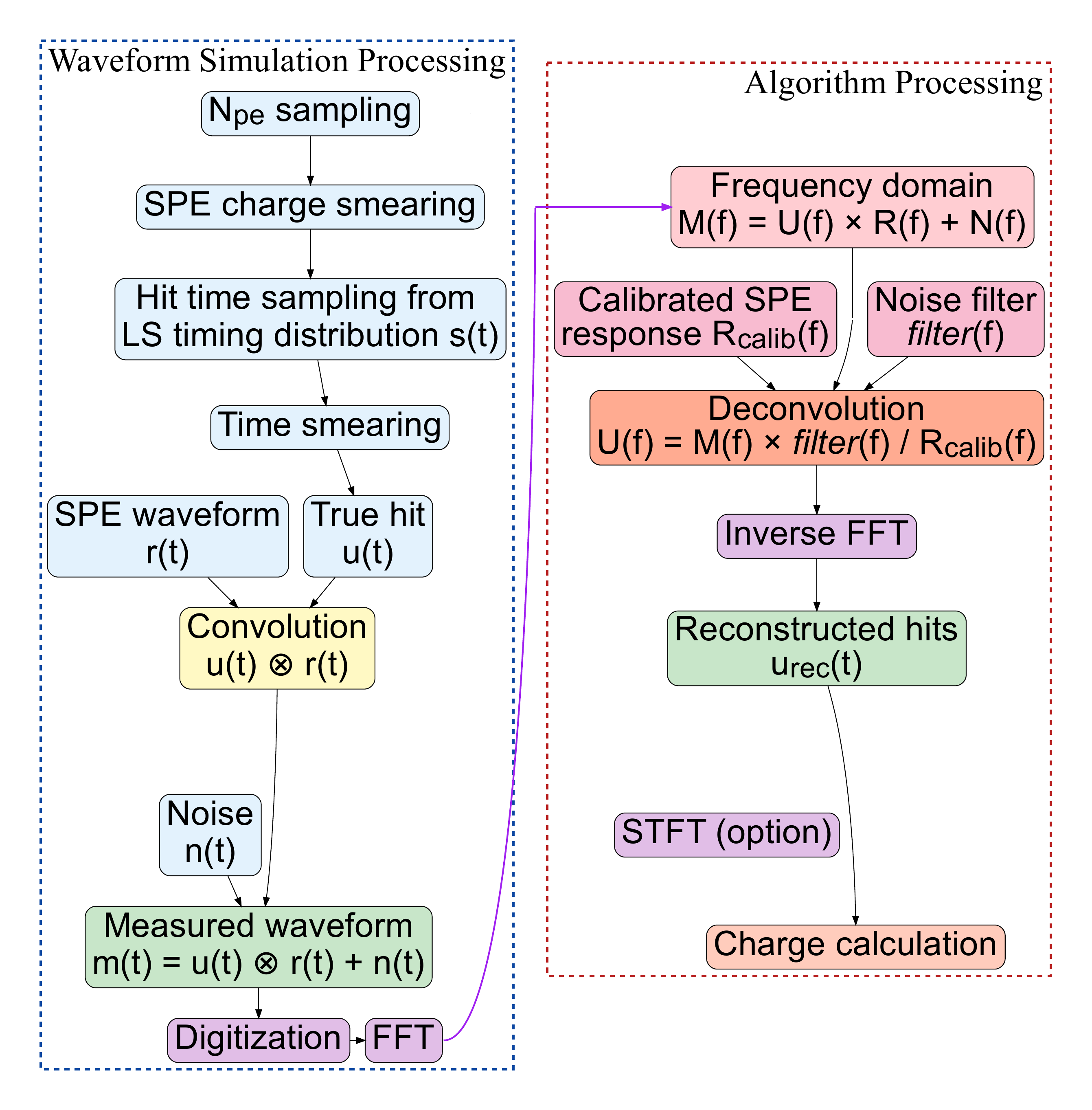}
\caption{\label{fig:pmt_decon_processing} The workflow for PMT waveform simulation  (left block diagram) and deconvolution-based reconstruction (right block diagram). In the simulation, true photoelectron hits $u(t)$ are convolved with the single photoelectron (SPE) response $r(t)$ and combined with Gaussian noise $n(t)$ to generate the measured waveform $m(t) = u(t) \otimes r(t) + n(t)$. The measured signal is transformed to the frequency domain via FFT. In the reconstruction, deconvolution is performed in the frequency domain: $U(f) = M(f) \times \mathit{filter}(f) / R_{\rm calib}(f)$, where $\mathit{filter}(f)$ suppresses noise and $R_{\rm calib}(f)$ is the calibrated SPE response. An inverse FFT transforms the result back to the time domain, yielding the reconstructed hits $u_{\text{rec}}(t)$, followed by charge calculation. The short-time Fourier transform may optionally be employed to perform time-frequency analysis, thereby enhancing the identification of pile-up hits.}
\end{figure}

The workflow for PMT waveform simulation is illustrated in Fig.~\ref{fig:pmt_decon_processing}. Each PMT waveform can be described as generated by convolving the photoelectron hit sequence \( u(t) \) with the SPE waveform, followed by the addition of electronic noise (modeled as white noise with a standard deviation of 0.1 PE in this study, referring to experimental measurements~\cite{Huang:2017abb,JUNO:2025fpc}). The photoelectron hit sequence \( u(t) \) encapsulates both the number and timing of photoelectron hits, and incorporates the charge and time dispersion effect. In the simulation, a Gaussian smearing with a standard deviation of 30\% is applied to the charge of each photoelectron, corresponding to the typical SPE charge resolution of 20-inch PMTs~\cite{Okajima:2015qoq,Xia:2015sfa,JUNO:2022hlz}. The temporal distribution of the photoelectron hit sequence \( u(t) \), also referred to as the time profile, depends on factors such as the liquid scintillator formulation and the type of incident particle. A detailed discussion of this aspect will be presented in Section~\ref{sec:performance}. In addition, timing dispersion effects caused by factors such as the PMT transit time spread (TTS) are also considered in the simulation, modeled by a Gaussian distribution with a standard deviation of 3~ns. To facilitate the calculation of the average baseline, a region of 120~ns is reserved before the main peak in all generated waveforms, effectively shifting the waveforms to the right by 120~ns. Subsequently, the simulated analog waveforms are digitized with a sampling rate of 1~GHz (sufficient to sample pulse with good fidelity) and a sampling length of 1000~ns (chosen to cover both fast and slow scintillation components, typically up to $\sim$200~ns).

\subsection{Review of the deconvolution algorithm}
\label{sec:deconvolution}

Deconvolution is a commonly used technique in digital signal processing~\cite{DSP}. The workflow of PMT waveform reconstruction based on deconvolution method is illustrated in Fig.~\ref{fig:pmt_decon_processing}. First, the raw PMT waveform is transformed into the frequency domain using the fast Fourier transform (FFT), and a low-pass filter is applied to suppress high-frequency noise. A deconvolution operation is then performed to eliminate shape features such as undershoot from the original waveform. The resulting signal is transformed back to the time domain via the inverse fast Fourier transform (IFFT), yielding the reconstructed photoelectron hit sequence \( u_{\text{rec}}(t) \). By integrating this sequence over an appropriate time window, the integrated area corresponds to the reconstructed number of photoelectrons (i.e., the charge). While the filter suppresses high-frequency noise, it also affects the shape and time spread of \( u_{\text{rec}}(t) \). To improve the identification of pile-up hits, the short-time Fourier transform (STFT) can be further introduced; more details can be found in~\cite{Tang:2024jfs}. The filter parameters and other reconstruction algorithm configurations used in this paper are consistent with those in~\cite{Tang:2024jfs}. For clarity and ease of presentation, reconstruction results that undergo additional STFT processing are denoted as "Rec. FFT+STFT" in the following sections, while those without STFT are denoted as "Rec. FFT".

\section{Performance of deconvolution reconstruction method under large dynamic range of charge and different scintillation time profiles}
\label{sec:performance}

This section begins with a brief overview of the time response characteristics of liquid scintillator, followed by an investigation into the performance of the deconvolution-based waveform reconstruction algorithm. The analysis focuses on two aspects: the adaptability of the algorithm across a large dynamic charge range, and its stability under varying scintillation time profiles. To this end, PMT waveform reconstruction is performed for both point-like and track-like energy deposition events (using through-going muons as a representative example for the latter).

\subsection{The time response characteristics of liquid scintillator}
\label{sec:time_response}

Liquid scintillator is a commonly used detection medium in nuclear and particle physics experiments. When an incident particle deposits energy in the scintillator, it induces ionization or excitation of the solvent molecules. Through a series of complex energy transfer processes~\cite{Zhang:2020mqz}, scintillation photons are ultimately produced. The scintillation emission process is not governed by a single time constant, but typically exhibits a composite decay consisting of multiple fast and slow components. Experimental studies have shown that the time response characteristics of liquid scintillator can be described by a multi-exponential model:

\begin{equation}
\label{eq:multiple_exp}
s \left(t \right) = \sum_{i=1}\frac{q_i}{\tau_i}e^{-\frac{t}{\tau_i}}
\end{equation}

where $\tau_i$ is the decay time constant of the $i^{th}$ scintillation component, and $q_i$ represents its relative fraction, with the sum over all components normalized to 1, i.e., $\textstyle\sum_{i}{q_i}=1$. Table~\ref{table:time properties} summarizes the time response parameters and corresponding liquid scintillator formulations reported in several experimental studies~\cite{MarrodanUndagoitia:2009kq,OKeeffe:2011dex,Elisei:1997tw}. Additionally, there are many other similar measurements~\cite{Li_2011,Steiger:2024nes}.

\begin{table}[tb]
  \caption{The time response parameters and corresponding liquid scintillator formulations reported in several experimental studies.}
  \label{table:time properties}
  \centering
  \normalsize
  \setlength{\tabcolsep}{2.5pt}
  \begin{tabular}{ccccccccccc}
    \hline
    \textbf{Solvent/solute (g/l)} & Excited by & \textbf{$\tau_1$} & \textbf{$\tau_2$} & \textbf{$\tau_3$} & \textbf{$\tau_4$} & \textbf{$q_1$} & \textbf{$q_2$} & \textbf{$q_3$} & \textbf{$q_4$} & Ref.\\
    \hline
    PXE/PPO(6.0) & $\gamma$ & 2.03 & 9.0 & 47.0 & 203.0 & 0.822 & 0.105 & 0.046 & 0.027 & \cite{MarrodanUndagoitia:2009kq} \\
    LAB/PPO(1.0) & $\gamma$ & 7.46 & 22.3 & 115.0 & - & 0.759 & 0.21 & 0.031 & - & \cite{MarrodanUndagoitia:2009kq} \\
    LAB/PPO(10.0) & $\gamma$ & 1.94 & 5.9 & 26.9 & 137.0 & 0.56 & 0.27 & 0.133 & 0.037 & \cite{MarrodanUndagoitia:2009kq} \\
    LAB(deoxy.)/PPO(2.0) & $\alpha$ & 3.2 & 18.0 & 190.0 & - & 0.44 & 0.16 & 0.41 & - & \cite{OKeeffe:2011dex} \\
    LAB(oxy.)/PPO(2.0) & $e^-$ & 4.3 & 16.0 & 166.0 & - & 0.75 & 0.22 & 0.03 & - & \cite{OKeeffe:2011dex} \\
    NE213 & $\alpha$ & 3.89 & 20.6 & 92.36 & 440 & 0.47 & 0.223 & 0.191 & 0.116 & \cite{Elisei:1997tw} \\
    PC/PPO(1.5) & $e^-$ & 3.57 & 17.61 & 59.5 & - & 0.895 & 0.063 & 0.042 & - & \cite{Elisei:1997tw} \\
    PC/PMP(6.0) & $e^-$ & 2.99 & 8.65 & 54.34 & - & 0.915 & 0.08 & 0.005 & - & \cite{Elisei:1997tw} \\
    \hline
  \end{tabular}
\end{table}

\begin{figure}[H]
\centering 
\includegraphics[width=0.7\textwidth]{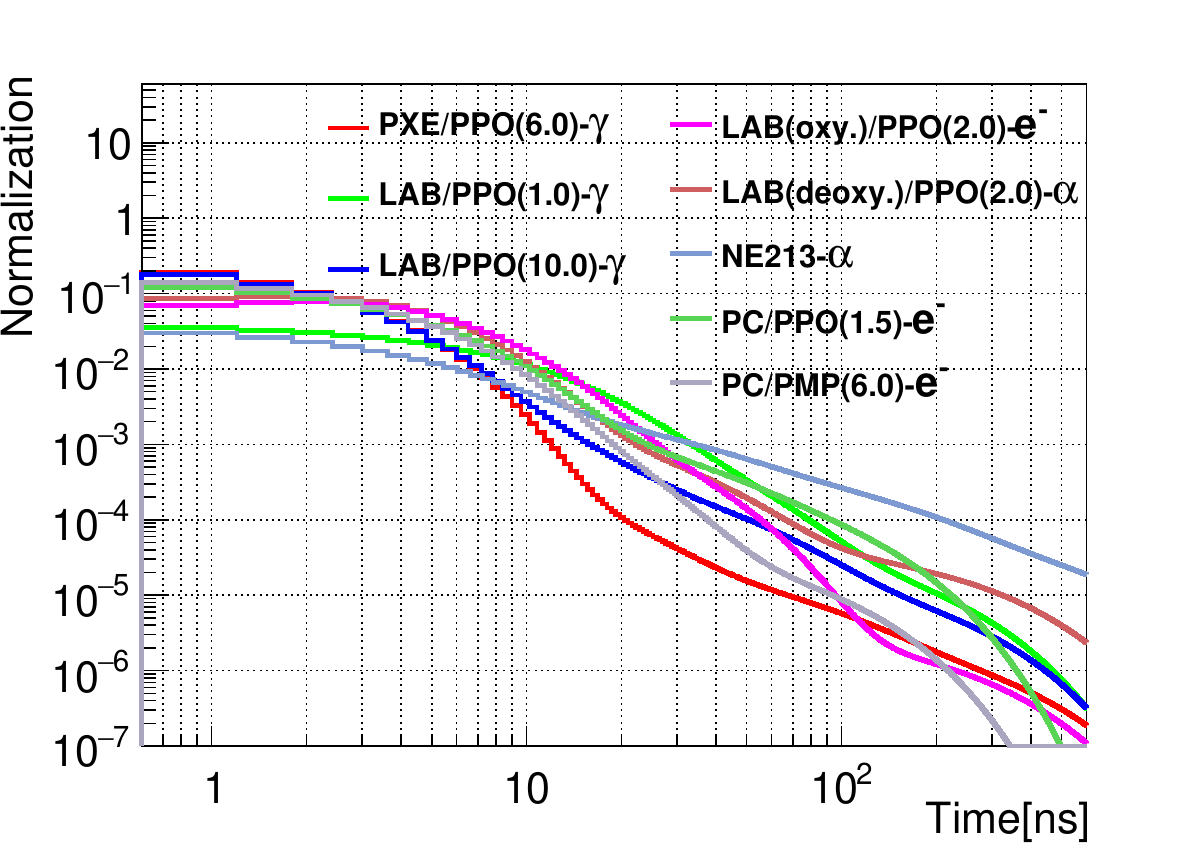}
\caption{\label{fig:pdf_selected}The scintillation time profiles for several representative liquid scintillator formulations listed in Table~\ref{table:time properties}.}
\end{figure} 

On the other hand, the scintillation time response of liquid scintillator exhibits particle dependence, meaning that the relative proportions of fast and slow components differ depending on the type of incident particle. In general, the scintillation decay processes induced by neutrons and alpha particles are slower than those induced by electrons and gamma rays. This characteristic forms the physical basis for particle identification using pulse shape discrimination (PSD) techniques~\cite{OKeeffe:2011dex,Ranucci:1994cs}. To illustrate this difference, Fig.~\ref{fig:pdf_selected} presents the scintillation time profiles calculated using Eq.~\ref{eq:multiple_exp} for several representative liquid scintillator formulations listed in Table~\ref{table:time properties}, under the particle excitation conditions employed in their respective original experiments. The results reveal clear distinctions among these profiles. Given that the reconstruction performance of PMT waveforms may be affected by the scintillation time profile of the liquid scintillator—particularly when complex features such as undershoot are present and become coupled with the timing characteristics—this issue warrants careful investigation. Accordingly, in the following sections, PMT simulated waveforms are generated using the several scintillation time profiles shown in Fig.~\ref{fig:pdf_selected} as inputs (as described in Section~\ref{sec:simulation}), and the reconstruction performance of the deconvolution algorithm is analyzed and compared.

\subsection{Reconstruction of PMT waveforms for point-like events}
\label{Reconstruction_point-like}

A point-like event refers to a scenario in which the particle deposits its energy within a small localized region in the liquid scintillator. In such events, the spatial dispersion of the energy deposition itself can be treated as a second-order effect and thus neglected; in other words, the event can be approximately regarded as an ideal point source. Consequently, for point-like events, the distribution of photon arrival times at each PMT can be considered as the intrinsic scintillation time profile of the liquid scintillator (Fig.~\ref{fig:pdf_selected}), further modified by an overall time shift arising from the variation in photon time-of-flight (TOF), as well as additional temporal dispersion ($\sigma$) introduced by the detection process—dominated primarily by the PMT transit time spread (TTS). This relationship can be described by the following expression:

\begin{equation}
\label{eq:point-like_LStimging}
s_{\rm point-like}\left(t \right) = Gaus\left(t-TOF, \sigma \right) \otimes s\left(t-TOF \right)
\end{equation}

\begin{figure}[H]
\centering 
\includegraphics[width=0.7\textwidth]{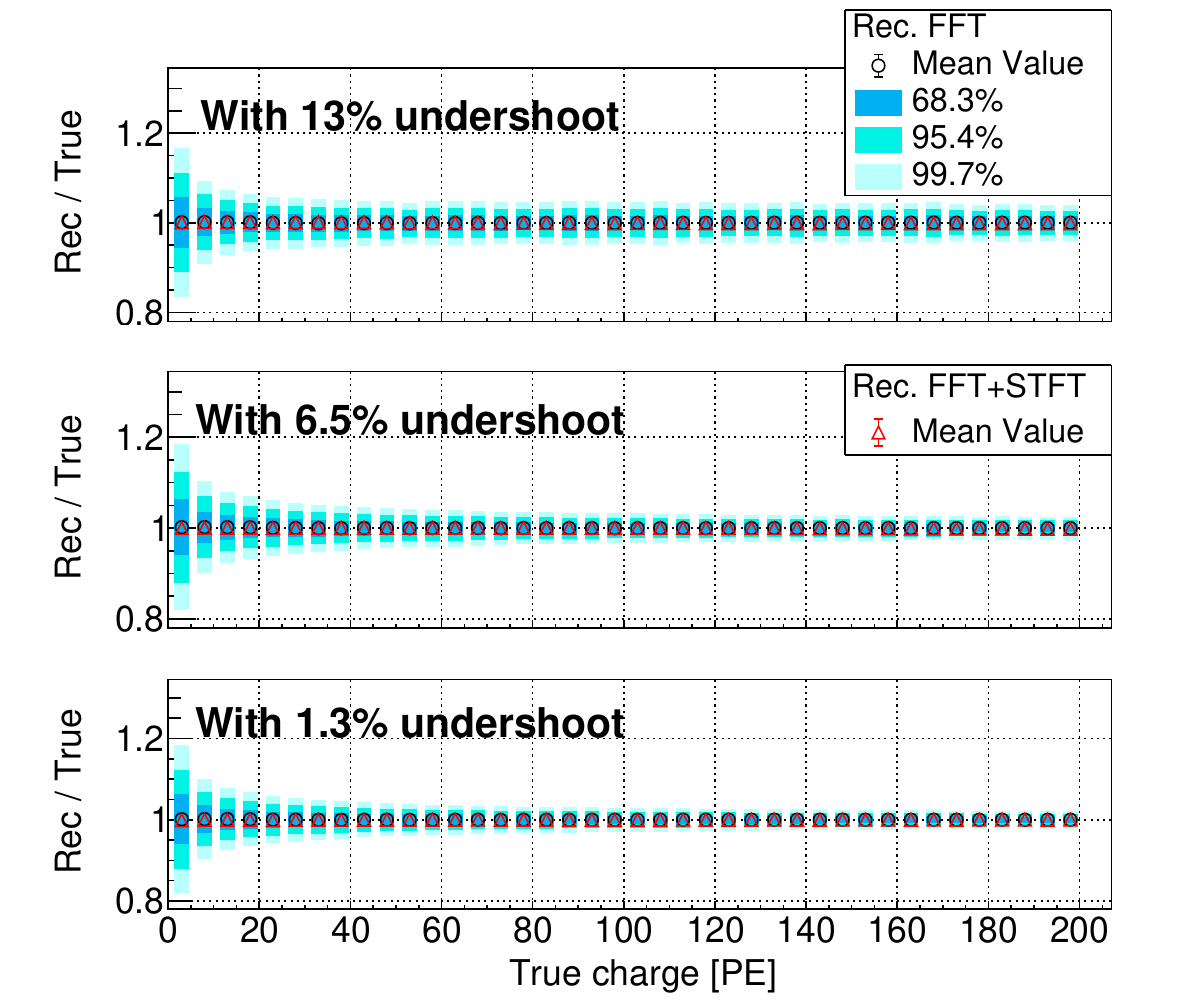}
\caption{\label{fig:compare_FFT_STFT_ovs_comp}Waveform reconstruction results of point-like events with different undershoot configurations. "Ratio" refers to the ratio of the reconstructed charge to the true charge. The three panels from top to bottom correspond to the undershoot configurations of 1.3\%, 6.5\%, and 13\% respectively. The reconstruction results obtained with the additional use of the short-time Fourier transform in the deconvolution-based reconstruction (denoted as "Rec. FFT+STFT") are consistent with those obtained without it (denoted as "Rec. FFT"). In each true charge bin, the black open circles and red open triangles represent the mean values of "Rec. FFT" and "Rec. FFT+STFT", respectively; their error bars are drawn but are too small to be visible in the figure. The three boxes in dark, medium and light shades of blue correspond to the 68.3\%, 95.4\%, and 99.7\% confidence intervals of the "Rec. FFT" reconstruction results. The distribution ranges of "Rec. FFT+STFT" at these three confidence intervals are very similar, but are not shown in the figure to avoid overcrowding; only its mean value is presented.}
\end{figure}

Using the first scintillation time profile listed in Table~\ref{table:time properties} (solvent: PXE, doped with 6 g/L PPO, excited by $\gamma$) as input, PMT waveform samples for point-like events were uniformly generated with charges distributed in the range of 0–200 PEs, following the simulation method described in Section~\ref{sec:simulation}. A total of 400,000 waveforms were produced and subsequently reconstructed using the deconvolution algorithm. The results are presented in Fig.~\ref{fig:compare_FFT_STFT_ovs_comp}. It can be observed that, under three different signal undershoot configurations, the deconvolution method achieves accurate reconstruction across the full dynamic range of 0–200 PEs, with residual non-linearity in the reconstructed charge below 1\% under the simulation conditions considered in this study. This dynamic range adequately covers the detection requirements for most physical targets in low- to medium-energy neutrino experiments. Additionally, it can be observed that when the undershoot is small, the dispersion of the reconstruction results is smaller.

\begin{figure}[H]
\centering 
\includegraphics[width=1.0\textwidth]{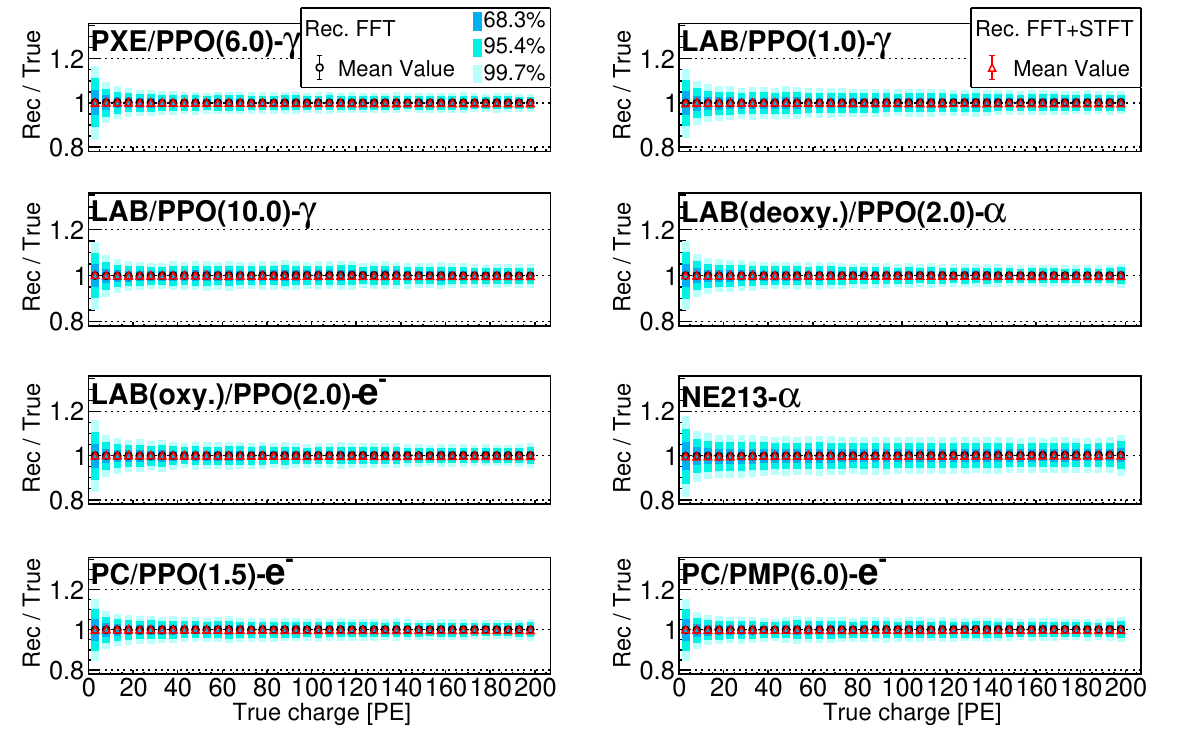}
\caption{\label{fig:compare_FFT_STFT_pdf_comp}Waveform reconstruction results of point-like events with different scintillation time profiles. "Ratio" refers to the ratio of the reconstructed charge to the true charge. The deconvolution algorithm demonstrates stable reconstruction performance, and the residual non-linearity of the reconstructed charge can be controlled within 1\%.}
\end{figure}

Furthermore, using the eight scintillation time profiles listed in Table~\ref{table:time properties} as inputs, we generated eight independent waveform datasets, each consisting of 400,000 waveforms covering a charge dynamic range of 0–200 PEs. For all waveforms, the undershoot of the SPE template is configured to be 13\% of the main peak amplitude. The reconstruction results (Fig.~\ref{fig:compare_FFT_STFT_pdf_comp}) demonstrate that the deconvolution algorithm exhibits stable performance across varying scintillation time profiles and a large charge dynamic range under the conditions of this simulation, confirming its adaptability in the aforementioned application scenarios.

\subsection{Reconstruction of PMT waveforms for track-like events}
\label{Reconstruction_track-like}

Track-like events are characterized by the highly extended spatial distribution of their energy deposition, as opposed to being confined to a small, localized region. In such events—exemplified by through-going muons—the emission sources of scintillation photons are distributed over a large volume. Consequently, the temporal characteristics of the PMT waveforms are influenced not only by the scintillation time profile and detector response but also by the geometric trajectory of the particle. Focusing on this feature, this section investigates the PMT waveform reconstruction for through-going muon events.

\begin{figure}[H]
    \centering
    \begin{subfigure}[t]{0.49\textwidth}
        \centering
        \includegraphics[width=\textwidth]{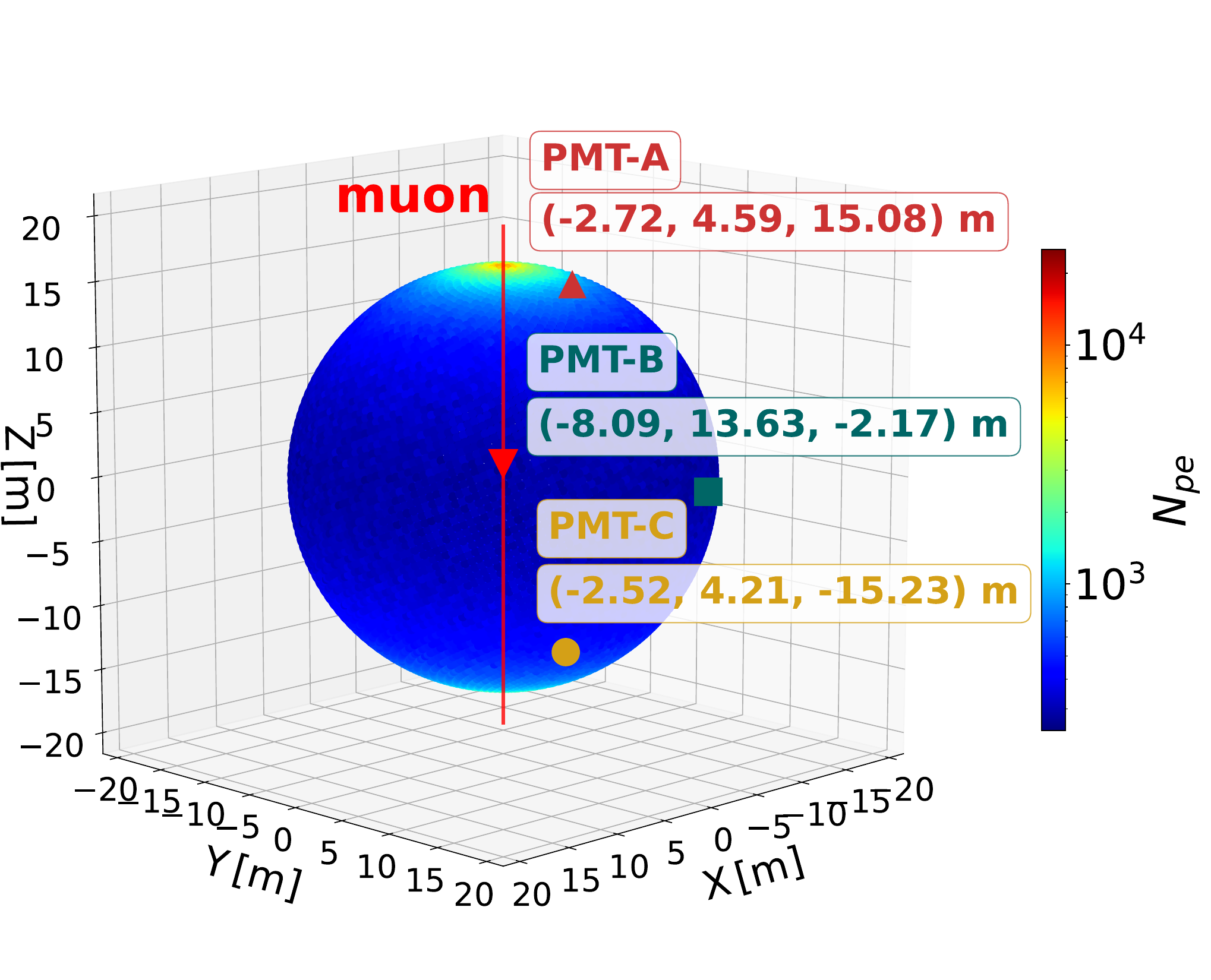}
        \caption{}
        \label{fig:pmt_distribution_with_ids}
    \end{subfigure}
    \hfill
    \begin{subfigure}[t]{0.49\textwidth}
        \centering
        \includegraphics[width=\textwidth]{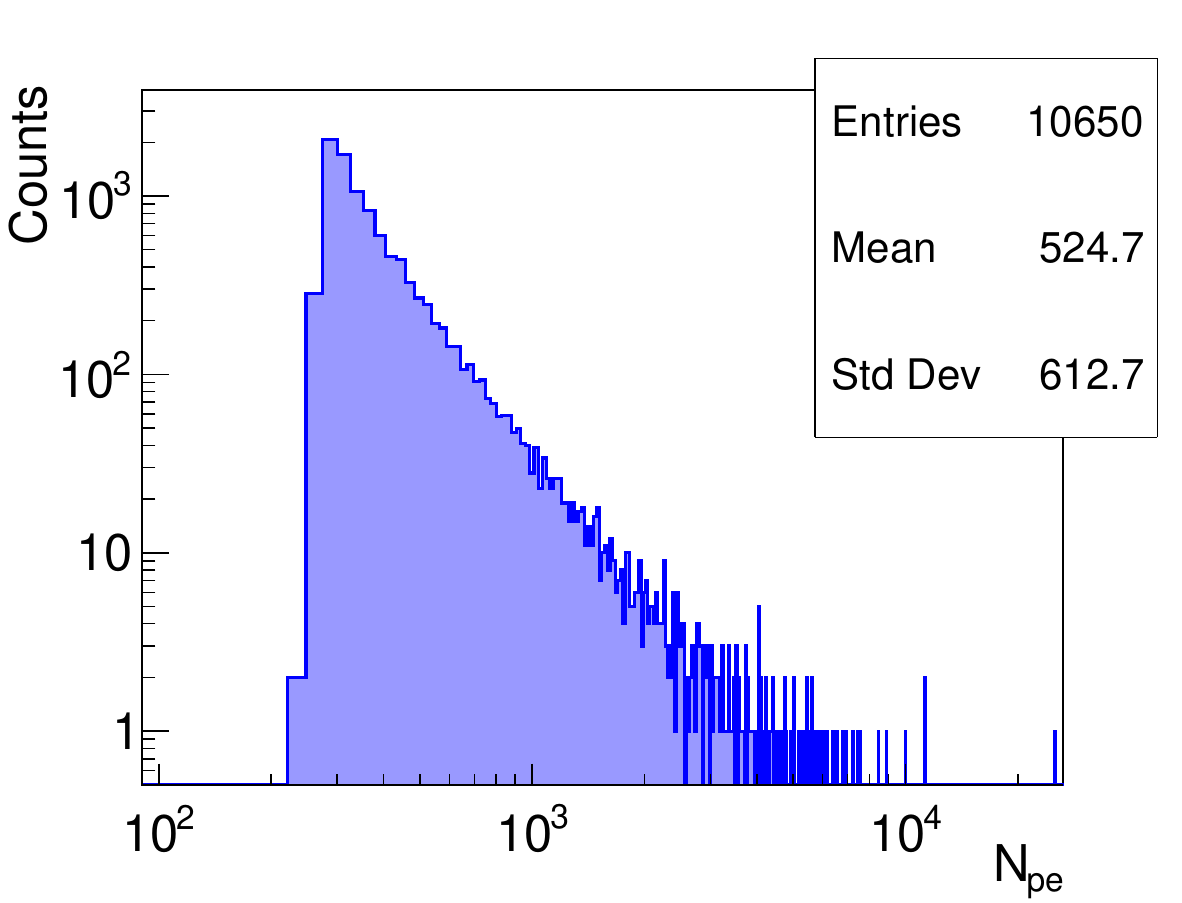}
        \caption{}
        \label{fig:npe_distribution}
    \end{subfigure}
    \caption{(a) Schematic diagram of the liquid scintillator detector for muon event simulation. The red track with an arrow is used to indicate the through-going muon. In the figure, the red triangle, green square, and yellow circle mark the positions of three PMTs located in the upper hemisphere, near the equator, and in the lower hemisphere, respectively. Their spatial coordinates (x, y, z) are (-2.72, 4.59, 15.08) m, (-8.09, 13.63, -2.17) m, and (-2.52, 4.21, -15.23) m, respectively. (b) The distribution of the number of photoelectrons detected by PMTs in a simulated muon event. The number of photoelectrons received by each PMT is mostly concentrated in the range of 300 to 500 PEs, while some PMTs receive up to several thousand PEs.}
    \label{fig:detector_simulation}
\end{figure}

First, a spherical liquid scintillator detector with a radius of 15~m was constructed using Geant4~\cite{GEANT4-2002zbu,Allison-2006ve,Allison-2016lfl}, as illustrated in Fig.~\ref{fig:pmt_distribution_with_ids}. The red track indicates the through-going muon, and the three markers indicate the positions of example PMTs. Fig.~\ref{fig:npe_distribution} presents the photoelectron distribution of the simulated muon event, which will be discussed later. In the simulation, the liquid scintillator is contained within a spherical acrylic vessel with a thickness of 10~cm. A total of 10,650 20-inch PMTs are employed for optical signal detection, positioned 1~m outside the acrylic vessel, with water filling the gap between the PMTs and the acrylic. This simulation is based on our previous work~\cite{Chen:2023xhj}. The optical properties of the liquid scintillator were configured following the parameters reported in~\cite{Zhou:2015gwa, Gao:2013pua, Wurm:2010ad, Ding:2015sys, Buck:2015jxa}, and a comprehensive set of optical processes was incorporated, including quenching, Rayleigh scattering, absorption, re-emission, photon propagation in the detector media, and reflection on acrylic surfaces. A muon with an energy of 200~GeV was simulated to propagate vertically downward from the north pole of the spherical detector, traversing the entire detector. This study only discusses through-going muons, i.e., those without high-energy electromagnetic or hadronic showers. The average energy deposition of the through-going muon in the liquid scintillator is approximately 2~MeV/cm. The deposited energy, distributed along a track length of about 30~m, is converted into scintillation photons, which then propagate through the detector and are ultimately detected by the PMTs. Photons incident on the PMT have a certain probability of being converted into photoelectrons, which ultimately contribute to the waveform output of the PMT. This probability is characterized by the PMT quantum efficiency, which is set to 30\% in this study.

Based on the Geant4 simulation described above, the number of photoelectrons ($N_{\rm PE}$) detected by each PMT and their corresponding arrival times can be obtained. By incorporating this information into the PMT waveform simulation, the waveform samples corresponding to muon events are subsequently generated. It is evident that the timing distribution of photoelectron hits on the PMT output waveform is inherently influenced by the relative geometric relationship between the energy deposition positions and the PMT. This relationship can be described by the following expression:

\begin{equation}
\label{eq:track-like_LStimging}
s_{\rm track-like}\left(t \right) = \sum_{j=1} w_j \times Gaus\left(t-TOF_j, \sigma \right) \otimes s\left(t-TOF_j \right)
\end{equation}

where $j$ indexes the $j$-th energy deposit, and $TOF_j$ is the time-of-flight for scintillation photons from the $j$-th energy deposit to reach the PMT under study. The term $w_j$ represents the weighting factor for energy deposit. It is worth noting that $w_j$ depends not only on the magnitude of the energy deposit but also on the attenuation of scintillation light, which is related to the relative geometric position between the energy deposit and the PMT. Eq.~\ref{eq:track-like_LStimging} serves as a conceptual illustration of the key factors influencing the timing distribution. In practice, the actual photoelectron hit times are obtained directly from Geant4 simulation and full optical propagation, which intrinsically account for all geometric and optical effects without relying on the simplified parameterization of Eq.~\ref{eq:track-like_LStimging}.

\begin{figure}[H]
\centering 
\includegraphics[width=0.5\textwidth]{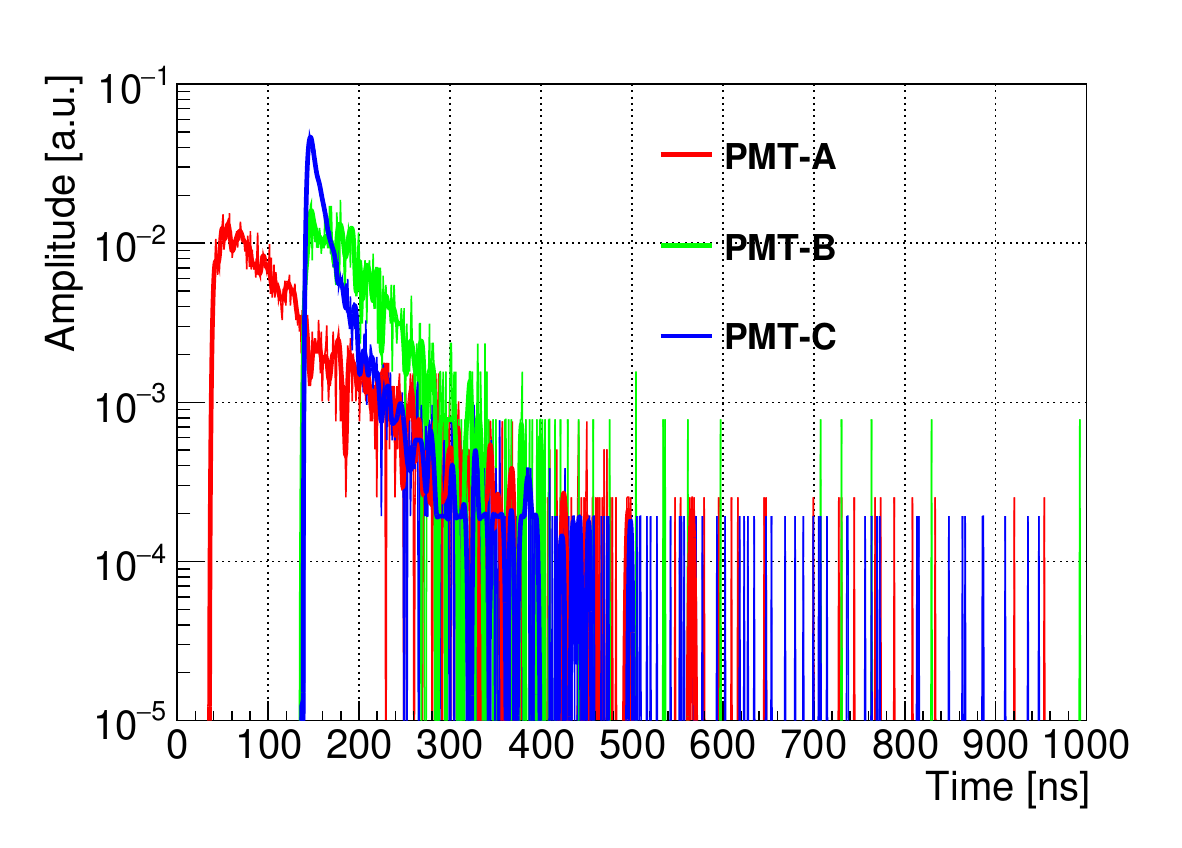}
\caption{\label{fig:Hittime_muon_file0}The photoelectron hit time distributions for three PMTs located at different positions. We use three PMTs (PMT-A, PMT-B and PMT-C) as an example: the PMTs at the upper hemisphere, the equatorial region, and the lower hemisphere of the detector (Fig.~\ref{fig:pmt_distribution_with_ids}).}
\end{figure} 

\begin{figure}[H]
    \centering
    \begin{subfigure}[t]{0.45\textwidth}
        \centering
        \includegraphics[width=\textwidth]{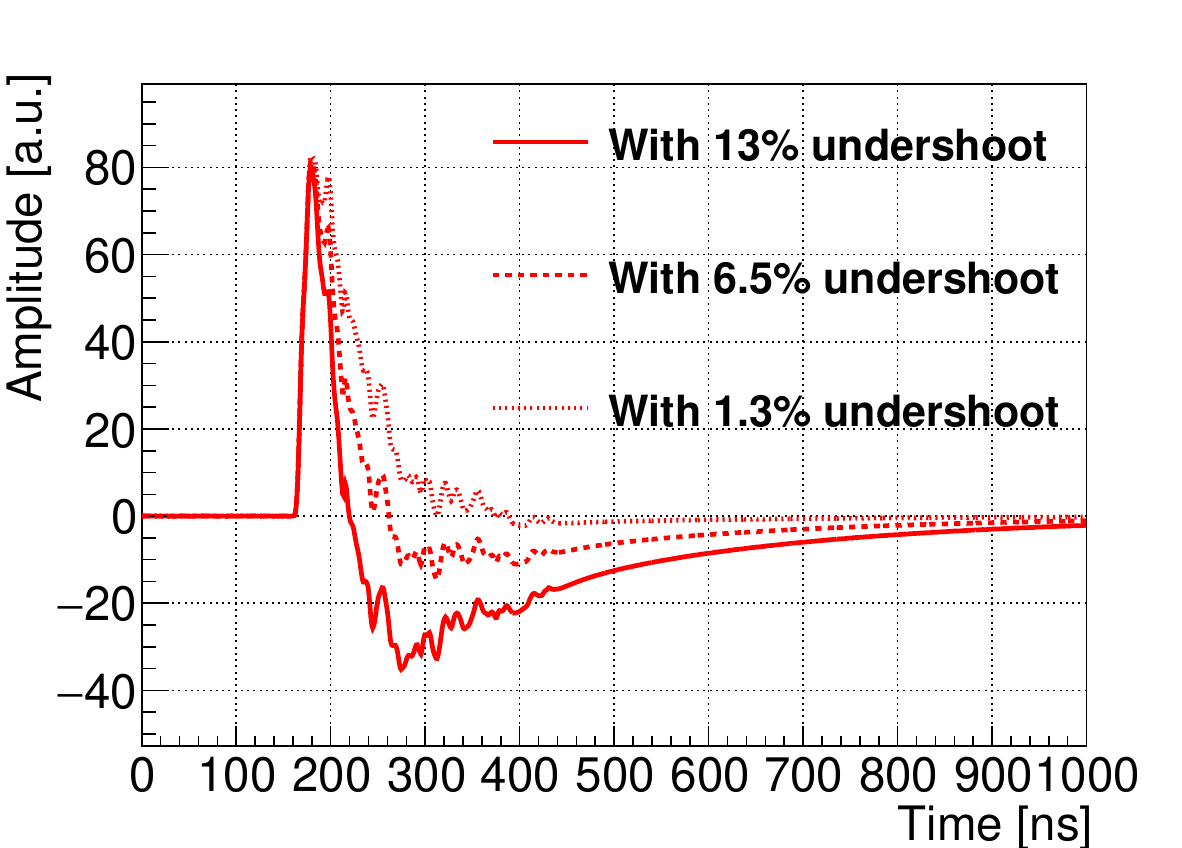}
        \caption{Waveform of PMT-A.}
        \label{fig:waveform_muon_raw_job_0_file_0_wf_0}
    \end{subfigure}
    \hfill
    \begin{subfigure}[t]{0.45\textwidth}
        \centering
        \includegraphics[width=\textwidth]{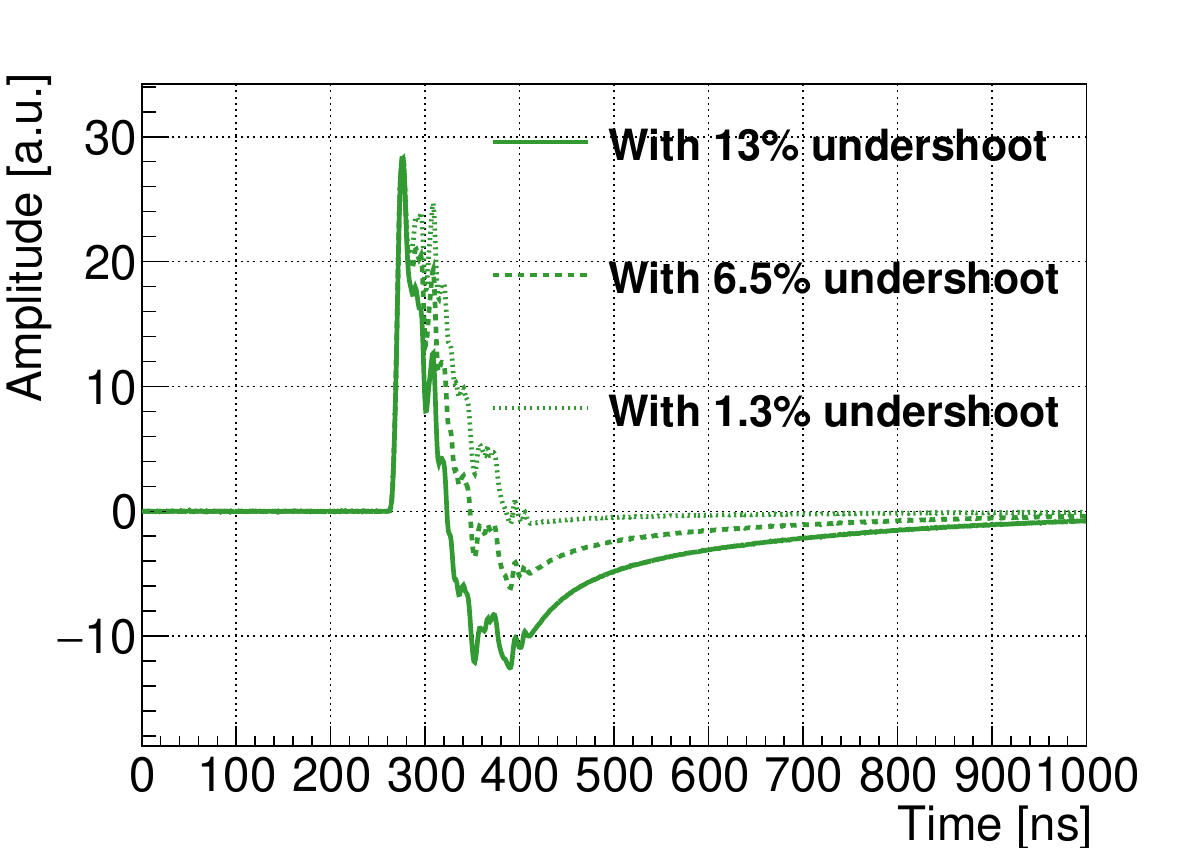}
        \caption{Waveform of PMT-B.}
        \label{fig:waveform_muon_raw_job_9_file_0_wf_900}
    \end{subfigure}
    \hfill
    \begin{subfigure}[t]{0.45\textwidth}
        \centering
        \includegraphics[width=\textwidth]{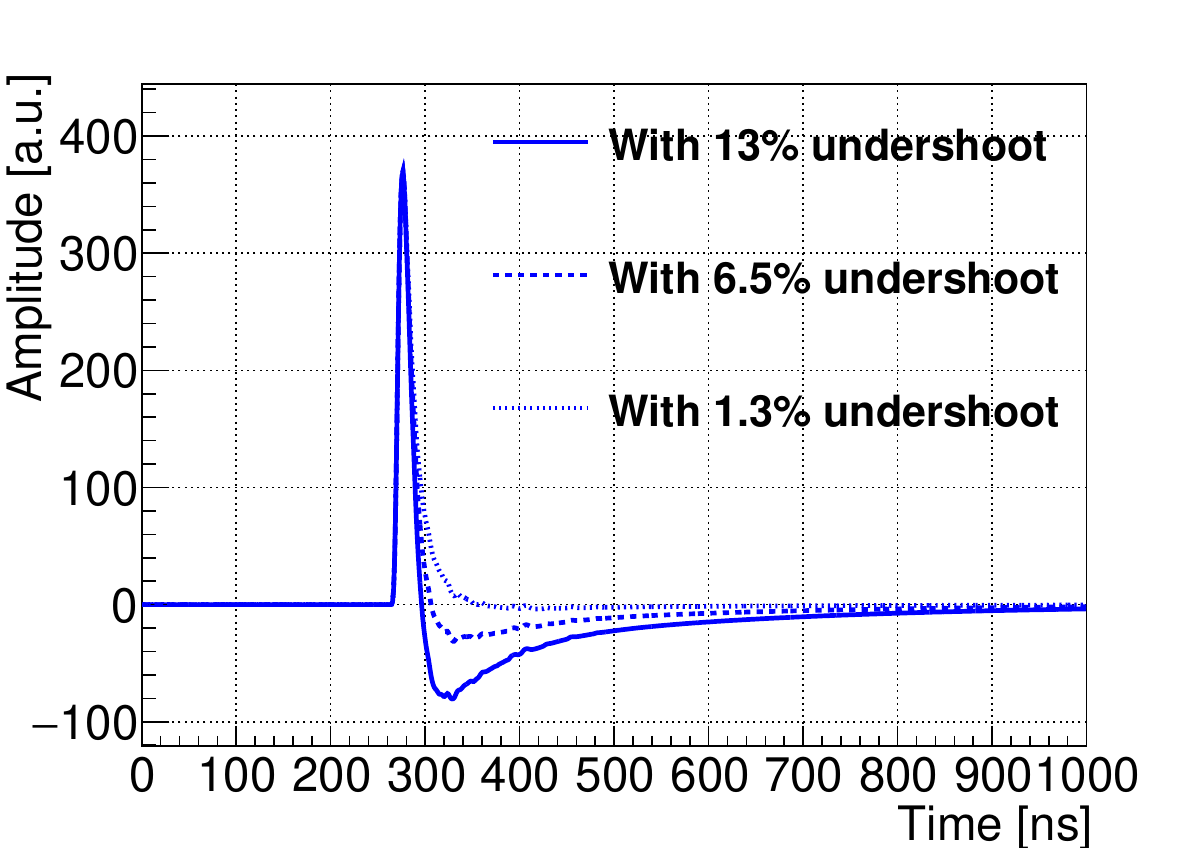}
        \caption{Waveform of PMT-C.}
        \label{fig:waveform_muon_raw_job_106_file_0_wf_10600}
    \end{subfigure}
    \caption{The simulated waveforms for PMT-A, PMT-B, and PMT-C are presented, corresponding to 1020~PEs, 286~PEs, and 1438~PEs, respectively. Waveforms for three different undershoot configurations are also shown.}
    \label{fig:muon waveform}
\end{figure}

As an example, Figure~\ref{fig:Hittime_muon_file0} shows the photoelectron hit time distributions for three PMTs located at different positions-corresponding to the upper hemisphere, the equatorial region, and the lower hemisphere of the detector. The positions of these three PMTs are indicated in Fig.~\ref{fig:pmt_distribution_with_ids} by a triangle, a square, and a circle, respectively. The simulation results indicate that PMT-A, being closest to the muon incident point, receives the optical signal earliest, while PMT-B and PMT-C exhibit significant delays in signal arrival, along with differences in the shape of their time distributions. As described by Eq.~\ref{eq:track-like_LStimging}, this behavior is primarily attributed to the relative positions between the PMTs and the energy deposition region of the muon in the liquid scintillator, as well as the optical propagation processes. The simulated waveforms corresponding to these three PMTs are shown in Fig.~\ref{fig:muon waveform}, with the numbers of photoelectrons ($N_{\rm PE}$) being 1020~PEs, 286~PEs, and 1438~PEs, respectively. In addition, three different undershoot configurations were considered in the simulation, and three corresponding sets of PMT waveform samples were generated for subsequent analysis. It can be observed that these PMT waveforms do not fully return to baseline by the end of the time window (i.e., at $t$ = 1000~ns), where the waveform amplitude deviates from 0 a.u. This effect becomes more pronounced under configurations with larger undershoots. We characterize the baseline recovery status by the amplitude of the PMT waveform deviation from the baseline at $t$ = 1000~ns (denoted as $Amplitude_{t=1000~ns}$).

\begin{figure}[H]
\centering
\includegraphics[width=0.5\textwidth]{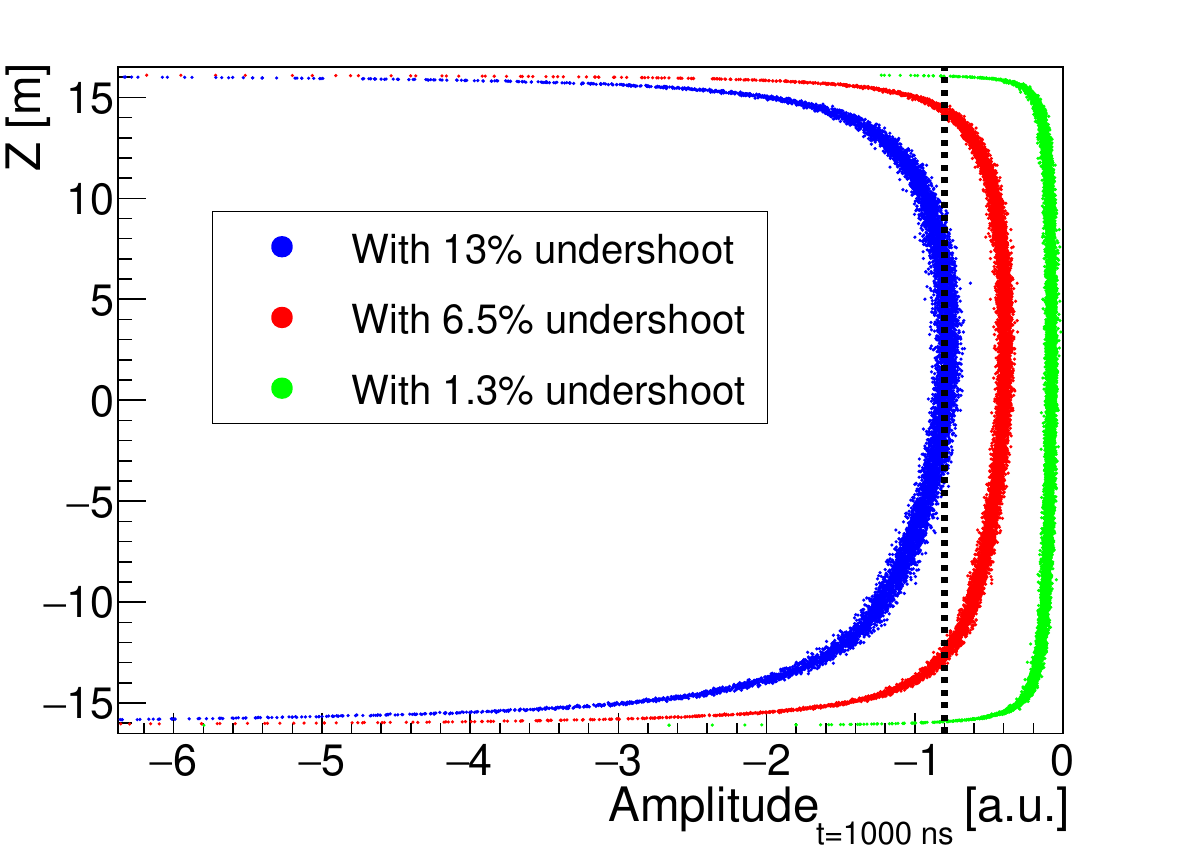}
\caption{\label{fig:tail_adc_vs_z} The baseline recovery status of PMTs located at different positions for a through-going muon traversing vertically downward through a spherical detector from its north pole. The green, red, and blue points correspond to configurations with SPE waveform undershoot set to 1.3\%, 6.5\%, and 13\%, respectively. The black dashed line corresponds to $Amplitude_{t=1000~ns} = 0.8~a.u.$ An examination shows that for SPE waveform undershoot configurations of 1.3\%, 6.5\%, and 13\%, the proportions of PMT waveforms with $Amplitude_{t=1000~ns} < -0.8~a.u.$ are approximately 0.7\%, 16.1\%, and 82.6\%, respectively.}
\end{figure}

Figure~\ref{fig:tail_adc_vs_z} illustrates the baseline recovery status for PMTs located at different positions. It can be observed that for a through-going muon event vertically incident from the top of the spherical liquid scintillator detector, PMTs near the north and south poles—being closer to the muon entry and exit points—detect a larger number of photoelectrons. Consequently, their waveforms do not fully return to the baseline by the end of the sampling window. Furthermore, when the undershoot amplitude of the SPE waveform is set to 13\%, the number of PMTs experiencing baseline recovery difficulties (with $Amplitude_{t=1000~ns}$ deviating significantly from zero) increases. Subsequent analysis establishes 
$Amplitude_{t=1000~ns} = 0.8~a.u.$ as an effective performance threshold for a 1000~ns PMT waveform: when $Amplitude_{t=1000~ns} < -0.8~a.u.$, the reconstruction quality degrades significantly. In the following, we present reconstruction results for two categories of waveforms: those meeting the criterion ($Amplitude_{t=1000~ns} \geq -0.8~a.u.$) and those that do not ($Amplitude_{t=1000~ns} < -0.8~a.u.$). A direct comparison between these two cases is provided to demonstrate the impact of baseline recovery on reconstruction performance.

\begin{figure}[H]
\centering 
\includegraphics[width=0.7\textwidth]{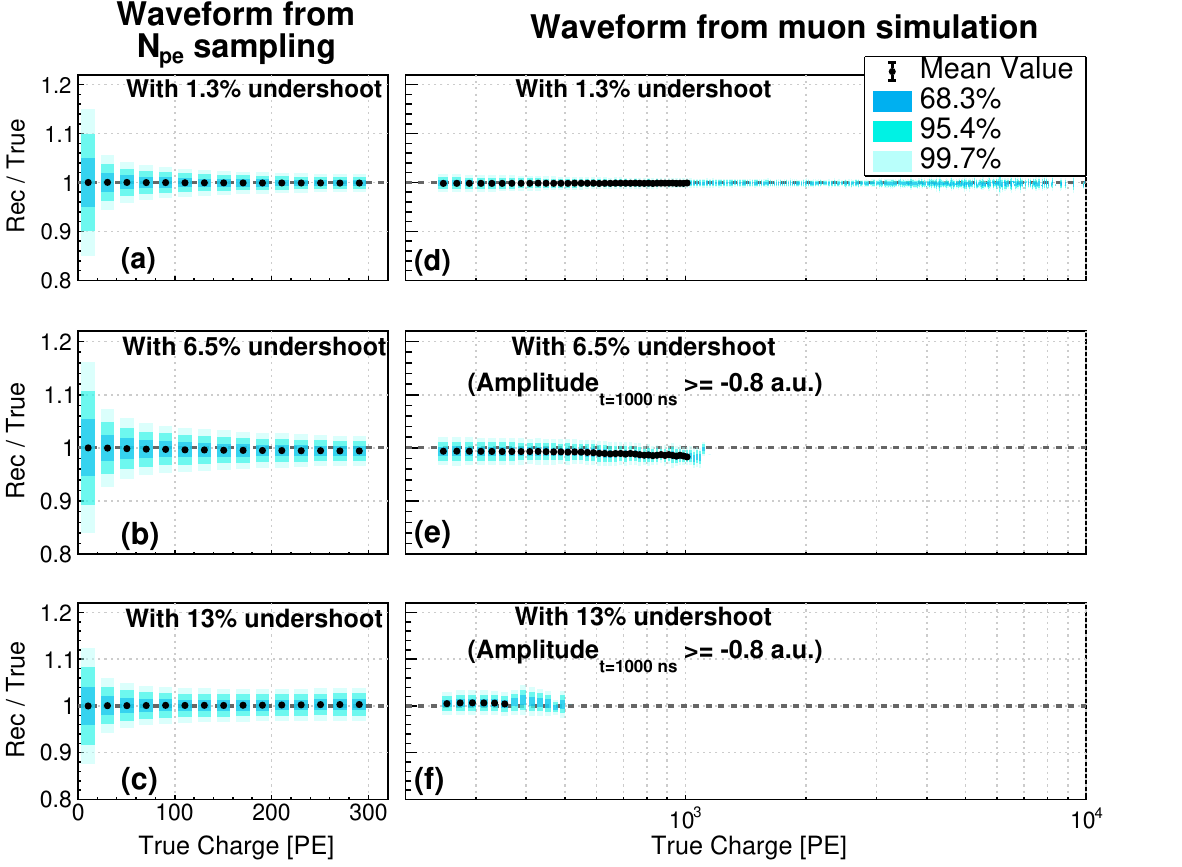}
\caption{\label{fig:ratio_comp_400pe_muonsim} The PMT waveform reconstruction performance for track-like events. The colors represent the number of waveforms in each bin, while the black points indicate the mean value of the ratio of reconstructed charge to true charge, with particular attention given to performance within the dynamic range up to 1000~PEs. The white dashed line corresponds to “${\rm Rec / True = 1}$", serving as a convenient reference for comparison. The three subfigures on the left correspond to simulated waveforms generated using the photoelectron hit time distributions of 10,650 PMTs obtained from muon simulations, with photoelectron numbers uniformly distributed between 0 and 300 PEs. These subfigures ((a), (b) and (c)) adopt SPE waveform undershoot amplitudes of 1.3\%, 6.5\% and 13\%, respectively. The three subfigures on the right ((d), (e) and (f)) correspond to PMT simulated waveforms from ten through-going muon events, with undershoot amplitudes of 1.3\%, 6.5\% and 13\%, respectively. An additional selection criterion is applied in subfigures (e) and (f), requiring that the baseline deviation amplitude of the PMT waveform at $t$ = 1000~ns ($Amplitude_{t=1000~ns}$) be greater than or equal to -0.8~a.u.}
\end{figure}

This section focuses on reconstructing large-signal PMT waveforms from through-going muon events. As simulation results (Fig.~\ref{fig:npe_distribution}) show each PMT receives over 200~PEs, identifying pile-up hits becomes impractical and offers marginal benefit to event reconstruction. Consequently, the STFT option is omitted from the reconstruction process in this section to reduce computational overhead. Figure~\ref{fig:ratio_comp_400pe_muonsim} presents the PMT waveform reconstruction performance for track-like events, based on two simulated waveform datasets:

(1) The first dataset consists of PMT simulated waveforms corresponding to ten through-going muon events, totaling 106,500 waveforms. These are shown in the three subfigures on the right side of Fig.~\ref{fig:ratio_comp_400pe_muonsim}, each corresponding to a different undershoot configuration. The number of photoelectrons for these waveforms ranges from 200 to 10,000 PEs, with the majority concentrated in the 250–500 PEs range.

(2) The second dataset is generated using the photoelectron hit time distributions of 10,650 PMTs obtained from the aforementioned muon simulation (three examples are shown in Fig.~\ref{fig:Hittime_muon_file0}) as input. Waveforms with photoelectron numbers uniformly distributed between 0 and 300 PEs are simulated, yielding a total of 106,500 waveforms. These are shown in the three subfigures on the left side of Fig.~\ref{fig:ratio_comp_400pe_muonsim}, also under different undershoot configurations. This dataset is primarily intended to supplement the first dataset by extending the coverage to photoelectron numbers below 200 PEs.

\begin{figure}[H]
    \centering
    \begin{subfigure}[t]{0.49\textwidth}
        \centering
        \includegraphics[width=\textwidth]{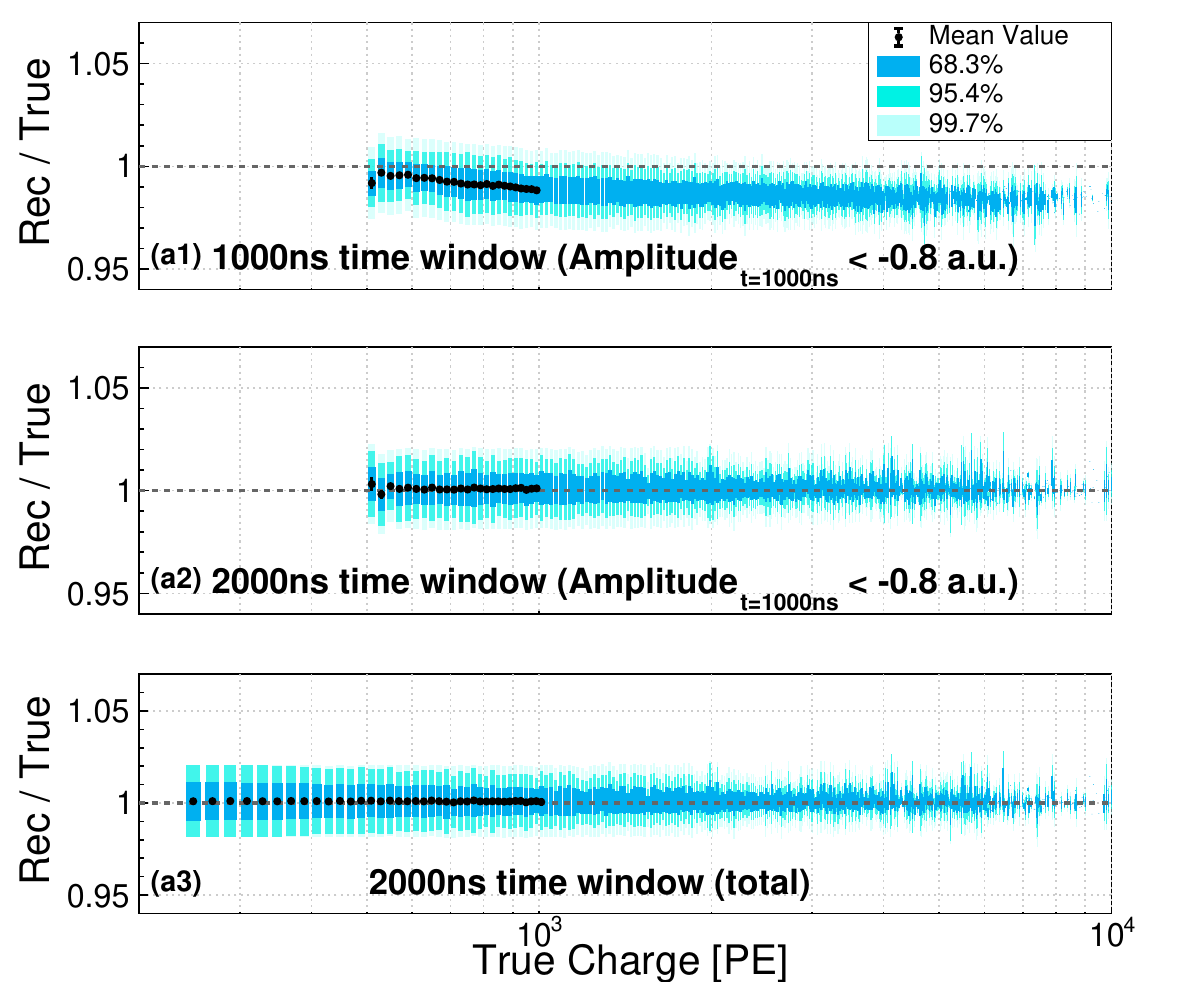}
        \caption{Reconstruction results for PMT waveforms with the configuration of 6.5\% undershoot.}
        \label{fig:ratio_vs_trueCharge_2000ns_comparison_A}
    \end{subfigure}
    \hfill
    \begin{subfigure}[t]{0.49\textwidth}
        \centering
        \includegraphics[width=\textwidth]{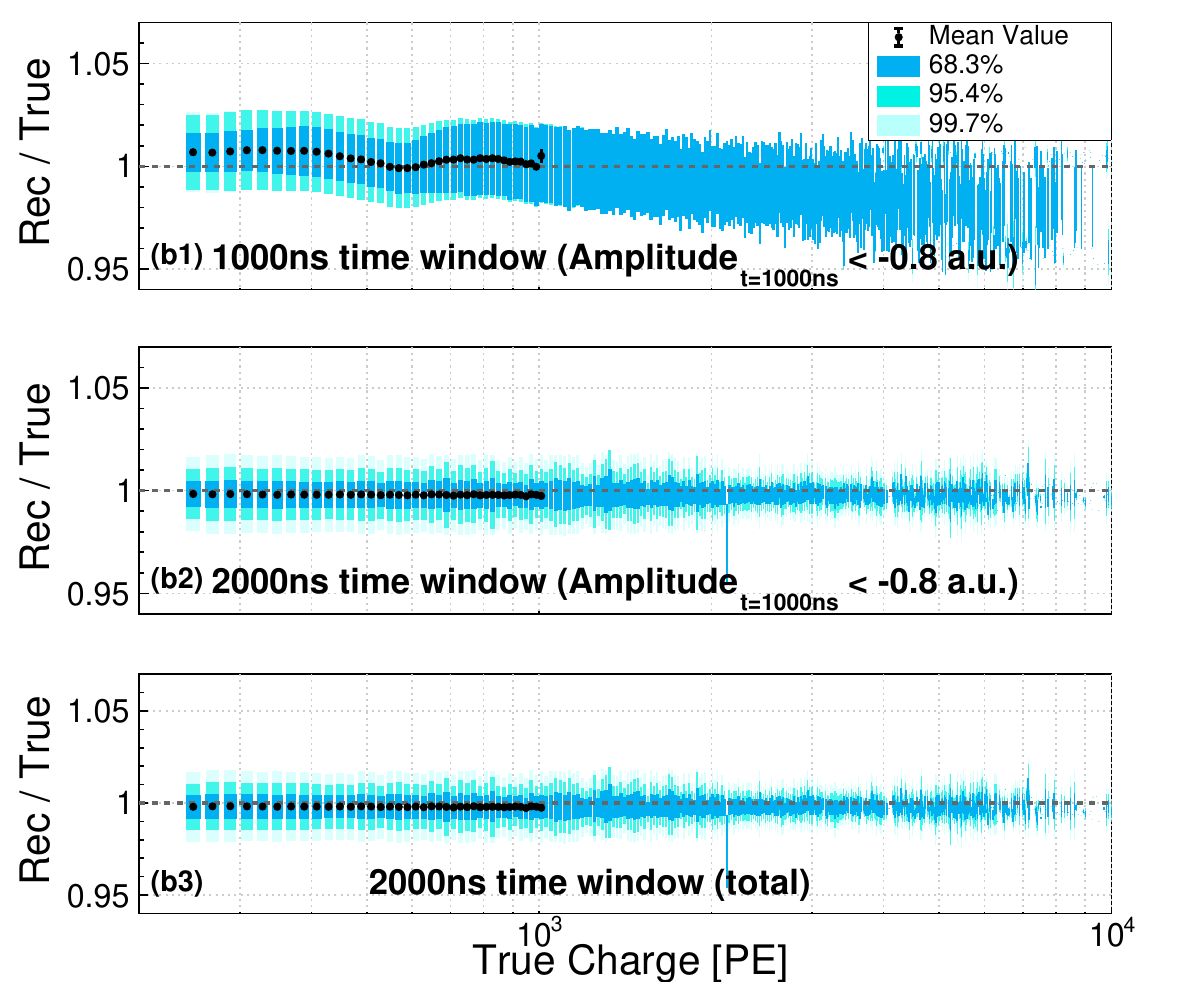}
        \caption{Reconstruction results for PMT waveforms with the configuration of 13\% undershoot.}
        \label{fig:ratio_vs_trueCharge_2000ns_comparison_B}
    \end{subfigure}
    \caption{Comparison of PMT waveform reconstruction performance for track-like events under different time windows. The colors represent the number of waveforms in each bin, while the black points indicate the mean value of the ratio of reconstructed charge to true charge, with particular attention given to performance within the dynamic range up to 1000~PEs. The white dashed line corresponds to “${\rm Rec / True = 1}$", serving as a convenient reference for comparison. (a) Configuration with a SPE waveform undershoot amplitude of 6.5\%; (b) Configuration with a SPE waveform undershoot amplitude of 13\%. The subfigures under each configuration illustrate the following: (a1/b1) Reconstruction results for waveforms with $Amplitude_{t=1000~ns} < -0.8~a.u.$ using a 1000~ns time window; (a2/b2) Reconstruction results for waveforms with $Amplitude_{t=1000~ns} < -0.8~a.u.$ using a 2000~ns time window; (a3/b3) Reconstruction results for all track-like event waveforms using a 2000~ns time window.}
    \label{fig:ratio_vs_trueCharge_2000ns_comparison}
\end{figure}

The reconstruction results for these two simulated waveform datasets are presented in Fig.~\ref{fig:ratio_comp_400pe_muonsim}. It can be observed that, within the dynamic range of 0 to 300~PEs, the residual non-linearity of the reconstructed charge remains below 1\% for all three undershoot configurations. For waveforms with larger photoelectron numbers obtained from the muon simulation, a considerable fraction exhibit incomplete baseline recovery within the 1000~ns sampling window (as shown in Fig.~\ref{fig:tail_adc_vs_z}). After selecting waveforms satisfying $Amplitude_{t=1000~ns} \geq -0.8~a.u.$, the residual non-linearity of the reconstructed charge for these waveforms can be controlled within 2\%. For the remaining waveforms with $Amplitude_{t=1000~ns} < -0.8~a.u.$, subfigures (a1) (with 6.5\% undershoot) and (b1) (with 13\% undershoot) in Fig.~\ref{fig:ratio_vs_trueCharge_2000ns_comparison} present the reconstruction results using a 1000~ns sampling window. The results indicate that for waveforms that do not return to baseline by the end of the sampling window, the deviation of the reconstructed charge from the true value is more pronounced. As shown in Fig.~\ref{fig:ratio_vs_trueCharge_2000ns_comparison}(b1), this effect is particularly evident when the SPE waveform undershoot is large. When the charge is greater than 500~PE, the candle plot in Fig.~\ref{fig:ratio_vs_trueCharge_2000ns_comparison}(b1) shows a significant broadening of the “Rec/True” ratio distribution, indicating large reconstruction dispersion for high-charge waveforms. To investigate the origin of this broadening, we further examined the two-dimensional distribution of “Rec/True” versus “True Charge”. This examination reveals that the broadened distribution actually separates into two distinct clusters, approximately located in the ranges of 1.01–1.05 and 0.95–1.01, respectively. Further analysis shows that these two clusters correspond to PMTs located around the muon entry point in the upper hemisphere and those around the muon exit point in the lower hemisphere. This observation is consistent with the expectation that PMTs near the muon exit point receive photoelectrons with relatively later arrival times, making their reconstructed charge more susceptible to incomplete baseline recovery.

Additionally, we further compare these results with those obtained using an extended sampling window of 2000~ns (subfigures (a2) and (b2) in Fig.~\ref{fig:ratio_vs_trueCharge_2000ns_comparison}). It is evident that, in cases where the PMT waveform exhibits large undershoot, appropriately extending the sampling window can effectively mitigate the impact of incomplete baseline recovery on waveform reconstruction. Within 1000~PEs, the reconstructed charge demonstrates stable performance, with residual non-linearity below 1\%. It should be noted that practical applications may involve more complex scenarios. On one hand, an excessively long sampling window increases the data volume; on the other hand, the actual baseline behavior may be influenced by additional factors such as PMT afterpulse and electronic stability. In such cases, dynamic baseline correction or extrapolation of the waveform tail during offline analysis may be more feasible solutions.

\section{Summary}
\label{sec:summary}

This paper investigates the performance of the deconvolution-based waveform reconstruction algorithm under conditions of large dynamic charge range and varying LS scintillation time profiles. By configuring three undershoot amplitudes for the SPE waveform and eight scintillation time profiles, PMT waveform simulation data were constructed to comprehensively evaluate the algorithm's performance in reconstructing point-like event waveforms within a dynamic range of 0–200~PEs. The results indicate that the residual non-linearity of the reconstructed charge can be controlled within 1\%. The algorithm exhibits good adaptability to the dynamic range and effectively handles waveforms with different undershoot levels, while its reconstruction performance shows no significant dependence on the scintillation time profile. It should be noted that the 1\% residual non-linearity reported in this paper is obtained under the specific simulation conditions described in Section~\ref{sec:simulation}. Other factors not considered in this study (such as different noise levels, PMT saturation, or alternative detector configurations) may affect reconstruction performance. The applicability of these results to other experimental settings should be verified on a case-by-case basis.

Furthermore, the application of the algorithm to large-signal waveforms induced by through-going muons is examined. The results demonstrate that the deconvolution algorithm generally maintains effective reconstruction capability. For waveforms that do not fully recover to the baseline by the end of the sampling window, a slight degradation in reconstruction performance is observed, which can be mitigated by appropriately extending the sampling window. In practical applications, factors such as data volume constraints, PMT afterpulse, and electronic stability—which may influence baseline behavior—should be carefully considered, and corresponding mitigation strategies should be explored and implemented.

In summary, this study validates the reliability and applicability of the deconvolution-based PMT waveform reconstruction method under large dynamic charge ranges and various scintillation time profiles, providing a valuable reference for subsequent waveform processing and physical analysis.

\acknowledgments

This work was supported by National Key R\&D Program of China No. 2023YFA1606103, the Bagui Young Scholars Program of Guangxi Zhuang Autonomous Region, and the Academic Newcomer Award Program of Guangxi University.

\bibliographystyle{unsrt}
\bibliography{ref}
\end{CJK}
\end{document}